\documentclass[preprints,article,accept,moreauthors,pdftex]{Definitions/mdpi} 

\firstpage{1} 
\pubvolume{1}
\issuenum{1}
\articlenumber{0}
\pubyear{2021}
\copyrightyear{2020}
\datereceived{} 
\dateaccepted{} 
\datepublished{} 
\hreflink{https://doi.org/} 

\usepackage{todonotes}

\Title{Two sides of the same coin: sterile neutrinos and dark radiation. Status and perspectives.}

\TitleCitation{Title}

\newcommand{\DNeff}{\Delta N_\mathrm{eff}}
\newcommand{\Neff}{N_\mathrm{eff}}
\newcommand{\summnu}{\sum m_\nu}

\newcommand{\nua}[1]{\ensuremath{\rlap{\kern-2.5pt\ensuremath{\overset{\scriptscriptstyle(-)}{\phantom{\nu}}}}{\ensuremath{{\nu}_{#1}}}}}
\newcommand{\dm}[1]{\ensuremath{\Delta m_{#1}^2}}
\newcommand{\Us}[1]{\ensuremath{U_{#1}}}
\newcommand{\Usq}[1]{\ensuremath{|U_{#1}|^2}}

\Author{Maria Archidiacono $^{1}$\orcidA{}, Stefano Gariazzo $^{2}$\orcidB{}}

\AuthorNames{Maria Archidiacono and Stefano Gariazzo}

\AuthorCitation{Archidiacono, M.; Gariazzo, S.}

\address{%
$^{1}$ \quad Dipartimento di Fisica, Universit\`a degli Studi di Milano, Via G.\ Celoria 16, 20133 Milano, Italy;
and Istituto Nazionale di Fisica Nucleare (INFN), Sezione di Milano, Via G.\ Celoria 16, 20133 Milano, Italy; maria.archidiacono@unimi.it \\
$^{2}$ \quad Istituto Nazionale di Fisica Nucleare (INFN), Sezione di Torino, Via P. Giuria 1, I-10125 Turin, Italy; gariazzo@to.infn.it}

\abstract{The presence of light sterile neutrinos is one of the unanswered questions of particle physics. The cosmological counterpart is represented by dark radiation, i.e.\ any form of radiation  present in the early Universe besides photons and standard (active) neutrinos. This short review provides a comprehensive overview of the two problems and of their connection. We review the status of neutrino oscillation anomalies, commenting on the most recent oscillation data and their mutual tensions, and we discuss the constraints from other terrestrial probes. We show the shortcomings of translating light sterile neutrinos in cosmology as additional thermalised relativistic species, produced by neutrino oscillations, and we detail alternative solutions, specifically focusing on neutrino non standard interactions, and on their link to the Hubble constant problem. The impact of a new force leading to dark radiation -- dark matter interactions is also discussed in the realm of new physics in the dark sector. 
}

\keyword{Dark radiation; Sterile neutrinos
} 

\begin{document}

\section{Introduction}

High-precision observations of the Cosmic Microwave Background (CMB) radiation realized by the Planck satellite~\cite{Planck:2018nkj} indicate that most of the energy density of the current universe
is provided by some unknown components, commonly denoted as dark matter and dark energy.
Dark matter represents approximately 26.8\% of the current universe content and it is non-relativistic (see e.g.~\cite{Bergstrom:2012fi,Bertone:2016nfn,deSwart:2017heh,Luminet:2021xug}),
while dark energy is phenomenologically very similar to a cosmological constant and it is responsible for the recent acceleration of the Universe expansion (see e.g.~\cite{Copeland:2006wr,Sola:2016zeg,Huterer:2017buf,Lahav:2017ojz,Novosyadlyj:2017cnk,ORaifeartaigh:2017yby}).
While we have solid evidence that dark matter and dark energy exist, although we ignore their nature,
they may not be the only non-standard ``dark'' components of the Universe.
Indeed, despite being subdominant nowadays, new relativistic species
could have affected the evolution of the universe in the past.
It is therefore worth exploring the phenomenology of additional relativistic components, coined ``dark radiation'', that would have an impact on the thermodynamics of the early Universe and on cosmological observables.

One possible way to constrain the existence of dark radiation is to put bounds on
the effective number of relativistic species, $\Neff$,
which measures how much radiation energy density comes from relativistic particles different from photons.
In the standard scenario, only
three neutrino families
contribute to $\Neff$,
and since their decoupling does not take place instantaneously,
they provide $\Neff=3.044$~\cite{Froustey:2020mcq,Akita:2020szl,Bennett:2020zkv} (see also~\cite{Hansen:2020vgm,Froustey:2021azz} for the effect of neutrino asymmetries).
Any additional contribution to $\Neff$ would be due to non-standard neutrino properties
(see e.g.~\cite{Luo:2020fdt,Du:2021idh,deSalas:2021aeh})
or to dark radiation.

Over the last decade, a variety of dark radiation candidates has been studied.
In this short review we will focus on one
of them, especially interesting
given our current knowledge of particle physics: light sterile neutrinos~\footnote{The case of axions or Axion-Like Particles (ALPs) will not be discussed in this review.
Indeed, axions od ALPs are more frequently considered as possible dark matter candidates \cite{Kawasaki:2013ae,Marsh:2015xka,Dentler:2021zij},
while they contribute to dark radiation only if produced in thermal processes
(see e.g.~\cite{Archidiacono:2013cha,Archidiacono:2015mda,DiValentino:2015wba,DiValentino:2015zta,DiValentino:2016ikp,Mazumdar:2016nzr,Carenza:2021ebx,Giare:2021cqr} and references therein).}.
Indeed light sterile neutrinos have been proposed for many years
as a possible solution of the Short Baseline anomalies \cite{Gariazzo:2015rra,Giunti:2019aiy,Diaz:2019fwt,Boser:2019rta,Dasgupta:2021ies,Gariazzo:2021wsx}.
We discuss them in details in section~\ref{sec:sterile},
firstly listing the constraints we obtain from terrestrial probes,
and secondly analysing the impact of sterile neutrinos on cosmology.
In section \ref{sec:NSI} we focus on neutrino non-standard interactions as a mean to accommodate sterile neutrinos in cosmology,
also in light of the infamous Hubble constant problem \cite{Riess:2021jrx,Riess:2020fzl}.
In section~\ref{sec:darksector}, then, we extend the discussion to the case of dark radiation as part of 
some new physics extension of the standard model
(see e.g.~\cite{Heo:2015kra,Ko:2016fcd,Bringmann:2018jpr,Masina:2020xhk,Krnjaic:2020znf,Giovanetti:2021izc,Aloni:2021eaq,Niedermann:2021ijp,Niedermann:2021vgd}),
which may also be connected to dark matter within a complex dark sector.
Finally, section~\ref{sec:conclusions} summarizes our conclusions.

\section{Light sterile neutrinos}
\label{sec:sterile}
\subsection{Status of oscillation searches}
In this section we briefly summarize the status of sterile neutrino searches at oscillation experiments.
For a more extensive review, see for example refs.~\cite{Gariazzo:2015rra,Giunti:2019aiy,Diaz:2019fwt,Boser:2019rta,Dasgupta:2021ies,Gariazzo:2021wsx}.
We will discuss the different probes according to the oscillation channel they involve:
electron (anti)neutrino disappearance, muon (anti)neutrino disappearance, and neutrino appearance.
\begin{itemize}
\item Electron (anti)neutrino disappearance ($\nua{e}$ DIS) includes probes at reactors and Gallium experiments.
The Reactor Antineutrino Anomaly (RAA) \cite{Mention:2011rk} has been discovered in 2011, and it is related to the fact that observed reactor rates are smaller than the predicted ones.
Neutrino disappearance due to active-sterile oscillations has been proposed as a possible explanation to the anomaly.
The current status of the RAA has been recently summarized in ref.~\cite{Giunti:2021kab},
where the authors have shown that the anomaly disappears if one considers recent calculations of the reactor antineutrino flux \cite{Estienne:2019ujo,Kopeikin:2021ugh}.
The indications in favor of active-sterile neutrino oscillations, however, are also investigated 
using ``model-independent'' studies, which observe the reactor antineutrino flux at different distances in order to constrain the effect of oscillations separately from that of the absolute flux normalization and shape.
The current combined preference in favor of a sterile neutrino from such probes is currently below $2\sigma$ for a mass splitting $\dm{41}=1.3$~eV$^2$ and mixing angle $\sin^2 2\theta_{ee} = 0.026$
(upper limit $\Usq{e4}\lesssim0.015$ at $3\sigma$) \cite{Giunti:2020uhv},
if one ignores the debated results by Neutrino-4~\cite{Serebrov:2020kmd} (see also \cite{Almazan:2020drb,Serebrov:2020yvp,Giunti:2021iti}).
Concerning Gallium, the original anomaly was discovered by Gallex and SAGE, and was quantified to be around $3\sigma$~\cite{Giunti:2010zu}.
The significance decreases a bit when computed according to more recent cross-section estimates \cite{Kostensalo:2019vmv}, but it is revived by the very recent BEST results \cite{Barinov:2021asz}.
Notice that the best-fit parameters preferred by the Gallium anomaly are in tension with the RAA best-fit \cite{Giunti:2021kab}.
\item Muon (anti)neutrino disappearance ($\nua{\mu}$ DIS)
probes involve atmospheric or accelerator neutrinos.
In the former category we mainly find IceCube \cite{IceCube:2016rnb,IceCube:2017ivd,IceCube:2020phf},
which in the most recent data release points out a weak preference in favor of active-sterile oscillations
over the standard three-neutrino case, with $\dm{41}\sim4.5$~eV$^2$.
Accelerator experiments include MINOS/MINOS+~\cite{MINOS:2016viw,MINOS:2017cae},
which provides the strongest bounds ($\Usq{\mu4}\lesssim10^{-2}$ at $3\sigma$)
within the $\nua{\mu}$ DIS channel for a wide range of mass splittings,
and NO$\nu$A, which recently published the first constraints on active-sterile mixing from a long-baseline (LBL) experiment \cite{NOvA:2021smv}.
Although current probes cannot reach the required precision, future LBL experiments will be crucial
to study the effect of the two additional Dirac CP phases associated to the $\Us{e4}$ and $\Us{\tau4}$ matrix elements \cite{Palazzo:2020tye}.
\item Appearance (APP) experiments, finally, test the presence of electron (anti)neutrinos in a beam of muon (anti)neutrinos produced at accelerators.
The first anomalous appearance of events was observed by LSND \cite{LSND:2001aii},
with a significance of approximately $4\sigma$.
The presence of an anomaly was later confirmed at almost $5\sigma$ by MiniBooNE \cite{MiniBooNE:2020pnu}, whose best-fit results are however in tension with those from ICARUS \cite{ICARUS:2013cwr} and OPERA \cite{OPERA:2018ksq}.
The combination of these experiments indicates a preferred region with effective mixing
angles $10^{-3}\lesssim\sin^22\theta_{e\mu}\lesssim10^{-2}$
and mass splittings $0.3\lesssim\dm{41}/\text{eV}^2\lesssim1.5$, at $3\sigma$.
The very recent results \cite{MicroBooNE:2021zai,MicroBooNE:2021rmx} published by MicroBooNE \cite{MicroBooNE:2016pwy},
however, exclude the presence of an electron neutrino excess at low energies,
by discarding the possibility that the MiniBooNE excess is due to single-photon events at 95\% CL \cite{MicroBooNE:2021zai},
while at the same time ruling out electrons as the sole source of the excess at more than 97\% confidence \cite{MicroBooNE:2021rmx}.
MicroBooNE also reports that its measurements
are inconsistent with a $\nu_e$ interpretation of the MiniBooNE excess,
that however remains unexplained.
%
\end{itemize}

A combination of the above mentioned probes is problematic because of the discrepancy between appearance (excluding MicroBooNE) and disappearance constraints,
which indicate very different regions for the effective mixing angle $\sin^22\theta_{e\mu}$:
it should be larger than approximately $10^{-3}$ for APP probes, but smaller than few $10^{-4}$ according to DIS constraints.
The discrepancy, called APP-DIS tension, was already strong few years ago \cite{Gariazzo:2017fdh,Dentler:2018sju,Diaz:2019fwt},
for this reason a global combination of APP and DIS probes is statistically meaningless because the datasets are in tension.
One possible explanation for the strong discrepancy is that LSND and MiniBooNE observed
a signal that is not related to sterile neutrinos,
so that their results must not be considered in the global analysis:
at that point, the APP-DIS tension disappears \cite{Dentler:2018sju}.
This thesis may be supported by the recent results published by MicroBooNE \cite{MicroBooNE:2021rmx}.

\subsection{Status from other terrestrial probes}
The existence of a light sterile neutrino would affect also different kinds of measurements,
for example determinations of the absolute neutrino mass scale which exploit the kinematics of $\beta$ decay~\cite{Shrock:1980vy}.
While the lightest neutrino mass generates a shift of the endpoint of the $\beta$ energy spectrum,
the presence of heavier neutrinos is visible through kinks in the energy spectrum.
The study of each kink gives access to information on the mass and mixing of the heavier neutrinos:
for the $i$th mass eigenstate, the kink is located
at a position equal to the endpoint energy minus the mass $m_i$,
and its amplitude is proportional to the mixing between the neutrino mass eigenstate and the electron flavor, \Usq{ei}.
Experiments such as KATRIN \cite{KATRIN:2020dpx}, besides probing the neutrino mass scale \cite{Aker:2021gma},
are therefore also capable of testing the mass and mixing of additional neutrino states.
The KATRIN collaboration, in particular, published constraints on the sterile neutrino parameters in~\cite{KATRIN:2020dpx}
(see also \cite{Giunti:2019fcj}),
where they show that no kink has been observed, and derive upper limits on the mass $m_4$ and mixing \Usq{ei}:
current 95\% CL limits constrain the effective mixing angle
$\sin^2(2\theta_{ee})\lesssim0.43$ at $\dm{41}=10$~eV$^2$
or
$\sin^2(2\theta_{ee})\lesssim0.06$ at $\dm{41}=400$~eV$^2$.
The final bounds by KATRIN are expected to either confirm or rule out the results from several oscillation probes,
as the projected final sensitivity covers part of the preferred regions by Neutrino-4 \cite{Serebrov:2020kmd},
by the reactor \cite{Giunti:2021kab} and the Gallium anomaly \cite{Barinov:2021asz}.

If neutrinos are Majorana particles, the existence of sterile neutrinos would also be observable
in neutrinoless double $\beta$ decay (electron creation) probes \cite{Giunti:2015kza,Giunti:2017doy,Huang:2019qvq}.
This is a very rare process, whose half-life is proportional to the inverse, squared, of the effective Majorana mass \cite{Dolinski:2019nrj}
$m_{\beta\beta}=\sum_i\Us{ei}m_i$.
Notice that in the previous formula there is no absolute value around \Us{ei}: this means that the Majorana phases enter the calculation.
Because of the Majorana phases, the contributions from the different mass eigenstates may cancel with one another, and even if neutrino is a Majorana particle,
the half-life of the process can be arbitrarily close to infinity.
The picture, however, is different if we consider only three or four neutrinos:
a detection of the Majorana mass in specific ranges may be a smoking gun that there are more than three neutrinos \cite{Giunti:2015kza,Giunti:2017doy}.
When using available upper limits on the effective Majorana mass to derive constraints on the mass and mixing of the fourth neutrino,
the extension of the allowed $2\sigma$ region is similar to that of $\beta$ decay probes~\cite{Hagstotz:2020ukm}.

\subsection{Cosmological constraints}
\label{sec:darkradiation}

The discovery of neutrino oscillation anomalies paved the way for the search of new relativistic particles in the cosmos. These new forms of radiation, coined ``dark radiation'', are characterised by an equivalent number of neutrinos $\Neff$:
\begin{equation}
    \rho_{\rm r}=\rho_\gamma \left[1+ \frac{7}{8} \left( \frac{T_\nu}{T_\gamma}\right)^{4} \Neff \right] \, ,
\end{equation}
where $\Neff=\Neff ^ {\rm SM}+ \DNeff$ with $\Neff ^ {\rm SM} = 3.044$ \cite{Akita:2020szl,Froustey:2020mcq,Bennett:2020zkv}, $\rho_{\rm r}$ and $\rho_\gamma$ are, respectively, the radiation and the photon energy density. The ratio between the neutrino temperature and the photon temperature $T_\nu/T_\gamma=\left(4/11\right)^{1/3}$ comes from entropy density conservation $g_{*,{\rm S}}T^3={\rm const.}$, with $g_{*,{\rm S}}$ number of degrees of freedom at temperature $T$.
The new radiation component $\DNeff$ can be due to any additional light species, e.g., besides light sterile neutrinos,
hot dark matter axions \cite{Archidiacono:2015mda, Archidiacono:2013cha} (for recent reviews on axions see~\cite{DiLuzio:2020wdo, Marsh:2015xka}),
dark photons \cite{Alonso-Alvarez:2021pgy},
or extra light bosons \cite{Buckley:2014hja, Feng:2008mu}.
The contribution of new relativistic species to $\Neff$ depends on their decoupling temperature $T_{\rm DR}$ and internal number of degrees of freedom $g_{*,{\rm DR}}$:
\begin{equation}
\DNeff= \frac{\rho_{\rm DR}}{\rho_\nu}
=
\frac{g_{*,{\rm DR}}}{7/4}\left(\frac{T_{\rm DR}}{T_\nu}\right)^4
\,.
\label{eq:Tdec}
\end{equation}

In the case of a sterile neutrino, it has been shown that its decoupling temperature is equal to that of active neutrinos \cite{Gariazzo:2019gyi}, but its internal number of degrees of freedom depends on the thermalisation process, which is driven by neutrino oscillations.
Its momentum distribution function, in particular, emerges from non-resonant processes
and it is therefore proportional to that of active neutrinos \cite{Dodelson:1993je},
with a coefficient equal to $\DNeff$.
In any case, the study of active-sterile neutrino oscillations in the early Universe shows that the values of the additional mixing angles $(\theta_{14},\,\theta_{24},\,\theta_{34})$ and mass squared difference ($\Delta m_{41}^2$) preferred by terrestrial experiments point towards a full thermalisation of the additional sterile species \cite{Hannestad:2012ky,Gariazzo:2019gyi}, thus implying $\DNeff\sim 1$ or $\Neff \sim 4$.

The hypothesis of a fully thermalised cosmological component associated to the neutrino oscillation anomalies was still viable as of early 2013 \cite{Calabrese:2013jyk, Riemer-Sorensen:2013iql, Archidiacono:2011gq}.
Planck results \cite{Planck:2018vyg}, however, showed that a fully-thermalised sterile neutrino is disfavoured by cosmology with high-statistical significance: the value obtained by fitting temperature (TT), E-mode polarization (EE), and their cross correlation (TE) to a $\Lambda$CDM model with varying $\Neff$ is $\Neff = 2.92^{+0.36}_{-0.37} $ (95\% CL, Planck TT,TE,EE + low E).
Cosmological probes, therefore, can be interpreted to put strong constraints on active-sterile neutrino oscillations in the early universe \cite{Hagstotz:2020ukm}.

The incompatibility of light sterile neutrinos with cosmological data is twofold: on the one hand, as we have just discussed, they imply a larger number of relativistic degrees of freedom ($\Neff \sim 4$), on the other hand they also require a value of the hot dark matter density of the Universe way beyond the current bounds.
Indeed, the CMB-only upper bound $\summnu <0.24$ (95\% CL, Planck TT,TE,EE + low E + lensing) becomes even more stringent with the inclusion of external data: Baryonic Acoustic Oscillations (BAO) lower the limit to $\summnu <0.12$ eV (95\% CL, Planck TT,TE,EE + low E + lensing + BAO) \cite{Planck:2018vyg}, Lyman-$\alpha$ data or new DR16 BAO further improve it to $\summnu <0.09$ eV (95\% CL, Planck TT,TE,EE + low E + lensing + BAO + Lyman-$\alpha$ \cite{Palanque-Delabrouille:2019iyz}, or Planck TT,TE,EE + low E + lensing + BAO DR16 \cite{DiValentino:2021hoh}).
Notice that, when combining Planck TT,TE,EE + low E + lensing with the most recent prior on the Hubble constant by the SH0ES team \cite{Riess:2021jrx}, the 95\% CL bound reaches the minimum value allowed in the neutrino mass Normal Hierarchy $\summnu <0.06$ eV, due to the well known anti-correlation between $\summnu$ and $H_0$ \cite{Archidiacono:2016lnv}.
However, this result should be interpreted cautiously for it is obtained by means of a joint fit of incompatible data.
Independently on the Hubble constant, the neutrino mass bounds summarized above indicate that eV sterile neutrinos violate not only the constraints on $\Neff$, but also the bounds on $\summnu$.

This double tension motivated the search for physics beyond the Standard Model that could accommodate light sterile neutrinos in cosmology, as we will discuss in the next Section~\ref{sec:NSI}.

\section{Neutrino non-standard interactions}
\label{sec:NSI}

One attempt to solve the $\Neff$ tension is by means of a primordial neutrino asymmetry \cite{Saviano:2013ktj, Mirizzi:2013kva}, which delays the active-sterile neutrino oscillations, and, thus, reduces the sterile neutrino contribution to $\Neff$.
The asymmetry required to obtain a value of $\Neff$ compatible with Planck constraints, however, is so large ($|L_\nu| \gtrsim {\cal O} (10^{-2})$) that it leads to distortions in the distribution function of active neutrinos, which modifies the Big Bang Nucleosynthesis (BBN) predictions of the abundances of the primordial elements.
Moreover, the lepton asymmetry does not prevent a partial thermalisation of sterile neutrinos after decoupling ($T_{\rm dec} \sim 1$~MeV), thus it cannot solve the tension with the $\summnu$ bounds.

A lower value of $\Neff$ can also be achieved in low reheating scenarios \cite{Gelmini:2004ah}. The drawback of this scenario is that the temperature required to efficiently suppress the sterile neutrino production is few MeV \cite{Gelmini:2008fq, deSalas:2015glj}, thus neutrinos do not have enough time to thermalise before decoupling, once again inducing relevant changes in BBN.

Recently, neutrino non-standard interactions (NSI) have proven to be a way to accommodate sterile neutrinos in cosmology. Non-standard interactions among standard active neutrinos were already studied to understand the free-streaming properties and in connection with their impact on the neutrino mass bounds \cite{Beacom:2004yd, Archidiacono:2013dua, Cyr-Racine:2013jua}, and recently to estimate their effect on the calculation of $\Neff$~\cite{deSalas:2021aeh}.
The extension to the sterile neutrino sector devises a new hidden interaction mediated by either a vector boson (see the seminal papers \cite{Hannestad:2013ana, Dasgupta:2013zpn}) or by a pseduoscalar (see the seminal paper \cite{Archidiacono:2014nda}).
The new interaction acts as an additional matter-like term in the quantum kinetic equations governing neutrino oscillations in the early Universe.
Thus, NSI suppress the mixing angle, delaying the sterile neutrino production.
The degree of thermalisation of sterile neutrinos depends on the efficiency of this mechanism, which in turns depends on the coupling constant ($G_{\rm eff}$ or $g$): the contribution of sterile neutrinos to $\Neff$ can be adjusted by tuning the coupling constant.
This early Universe phenomenology is common to both the vector boson model and the pseudoscalar model.
However, at late times, the two models differ: while in the case of a vector boson mediator the interaction strength decreases during the evolution of the Universe, much like standard weak interactions, in the case of a pseudoscalar mediator the interaction strength increases yielding a rich phenomenology at low redshift.
Hence, the interaction is still efficient when sterile neutrinos become non-relativistic ($z=z_{\rm nr}$), and sterile neutrinos with a mass much larger than the mediator annihilate into the pseudoscalars, (almost) disappearing from the cosmic neutrino background.
In this scenario, the energy density of sterile neutrinos does not enter the hot dark matter budget measured at $z<z_{\rm nr}$, and, as a consequence, it does not violate the cosmological bounds on $\summnu$.
On the contrary, the vector boson model cannot avoid the neutrino mass bounds because of the absence of this late time phenomenology \cite{Chu:2018gxk, Mirizzi:2014ama}.

Besides accommodating sterile neutrinos in cosmology, neutrino non-standard interactions mediated by a light (pseudo)scalar are also investigated as a way to solve the Hubble constant problem \cite{Archidiacono:2020yey,Brinckmann:2020bcn, Blinov:2020hmc,Das:2020xke,Ghosh:2019tab,Kreisch:2019yzn,Blinov:2019gcj,Archidiacono:2016kkh} \footnote{For recent reviews on the Hubble constant problem and its solutions see~\cite{Freedman:2021ahq, Khalifeh:2021jfl, Shah:2021onj, Efstathiou:2021ocp, Schoneberg:2021qvd, Dainotti:2021pqg, DiValentino:2021izs, Riess:2019qba, Verde:2019ivm, Knox:2019rjx}.}.
\begin{figure}[!h]
\centering
\includegraphics[trim=0.5cm 0 0 0, width=0.5\textwidth, clip]{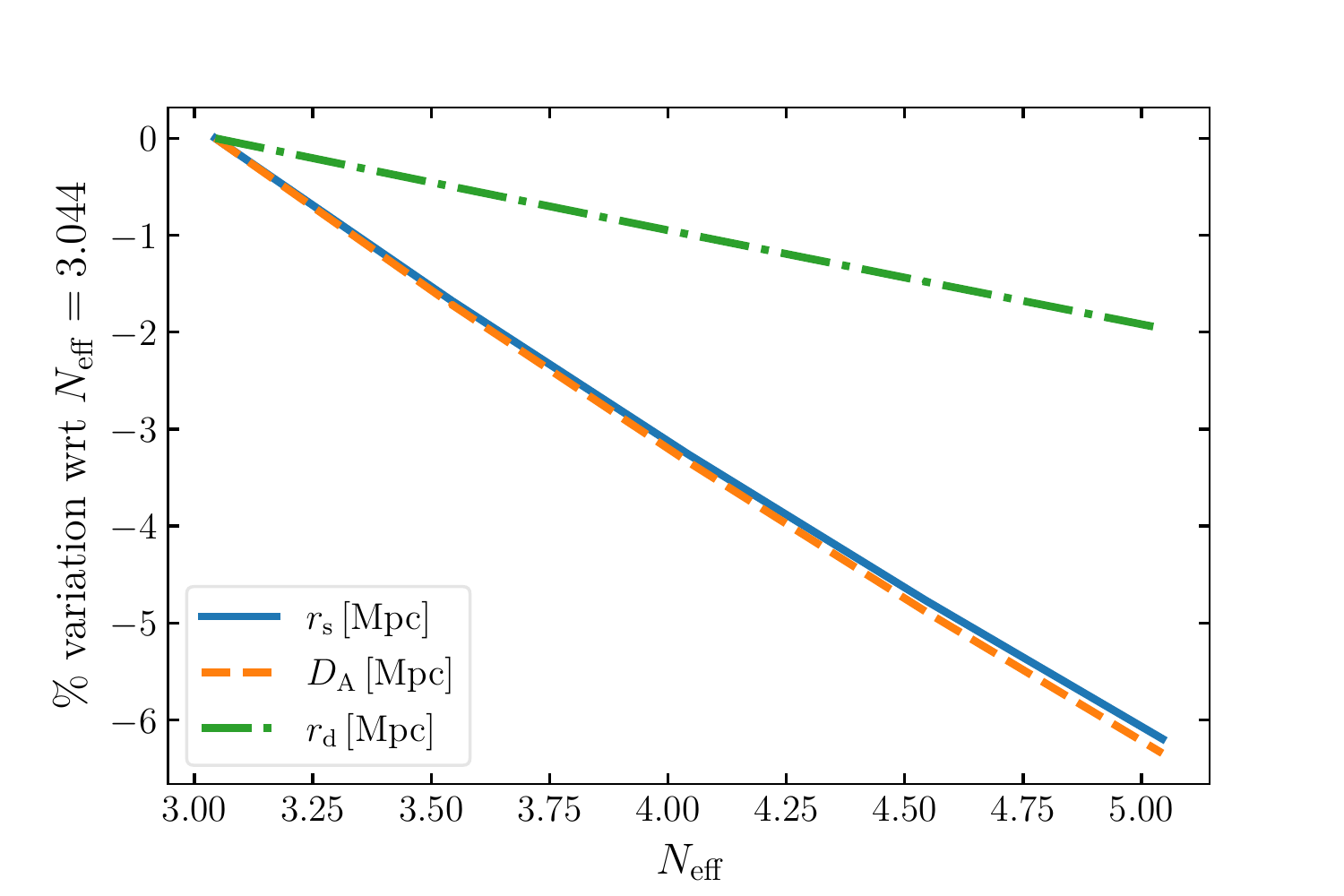}
\caption{Percentage variation of the distances $r_{\rm s}$ (blue solid line), $D_{\rm A}$ (orange dashed line), and $r_{\rm d}$ (green dot-dashed line) at recombination with respect to their reference values for $\Neff=3.044$. The distances are computed fixing $\theta_{\rm s}$, therefore $r_{\rm s}$ and $D_{\rm A}$ show similar variations.
}
\label{fig:dist}
\end{figure}
\begin{figure}[!h]
\includegraphics[trim=2cm 3.8cm 0 3.8cm, width=0.8\textwidth, clip]{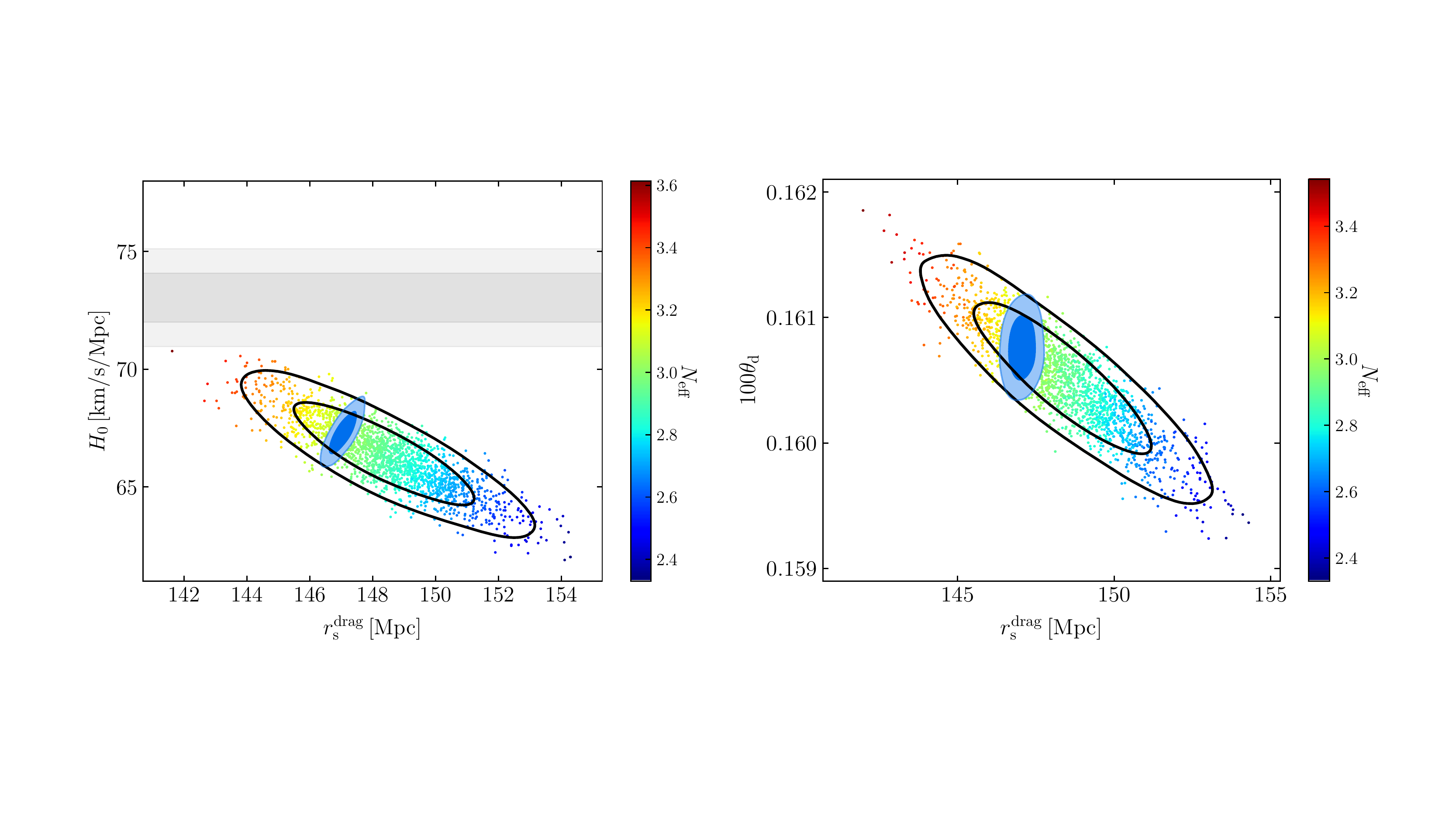}
\caption{
Marginalized 68\% CL and 95\% CL contours in the $H_0$-$r_{\rm s}^{\rm drag}$ plane (left panel) and in the $\theta_{\rm d}$-$r_{\rm s}^{\rm drag}$ plane (right panel) obtained by fitting Planck TT,TE,EE + low E to a pure $\Lambda$CDM model (blue filled contours), and to a model with varying $\Neff$ (black empty contours with points coloured according to the value of $\Neff$).
Larger values of $\Neff$ get closer to the $H_0$ value preferred by SH0ES \cite{Riess:2021jrx,Riess:2020fzl} (gray band) (left panel). However, this implies a smaller value of $r_{\rm s}$, which is correlated with a larger value of $\theta_{\rm d}$ (right panel). Note that here the comoving sound horizon is computed at the end of baryon drag, which is slightly after recombination; therefore $r_{\rm s}^{\rm drag}$ is slightly larger than the $r_{\rm s}$ of Figure~\ref{fig:dist}.
}
\label{fig:3D}
\end{figure}

\begin{figure}[!h]
\centering
\includegraphics[trim=0.4cm 0 0 0, width=0.5\textwidth, clip]{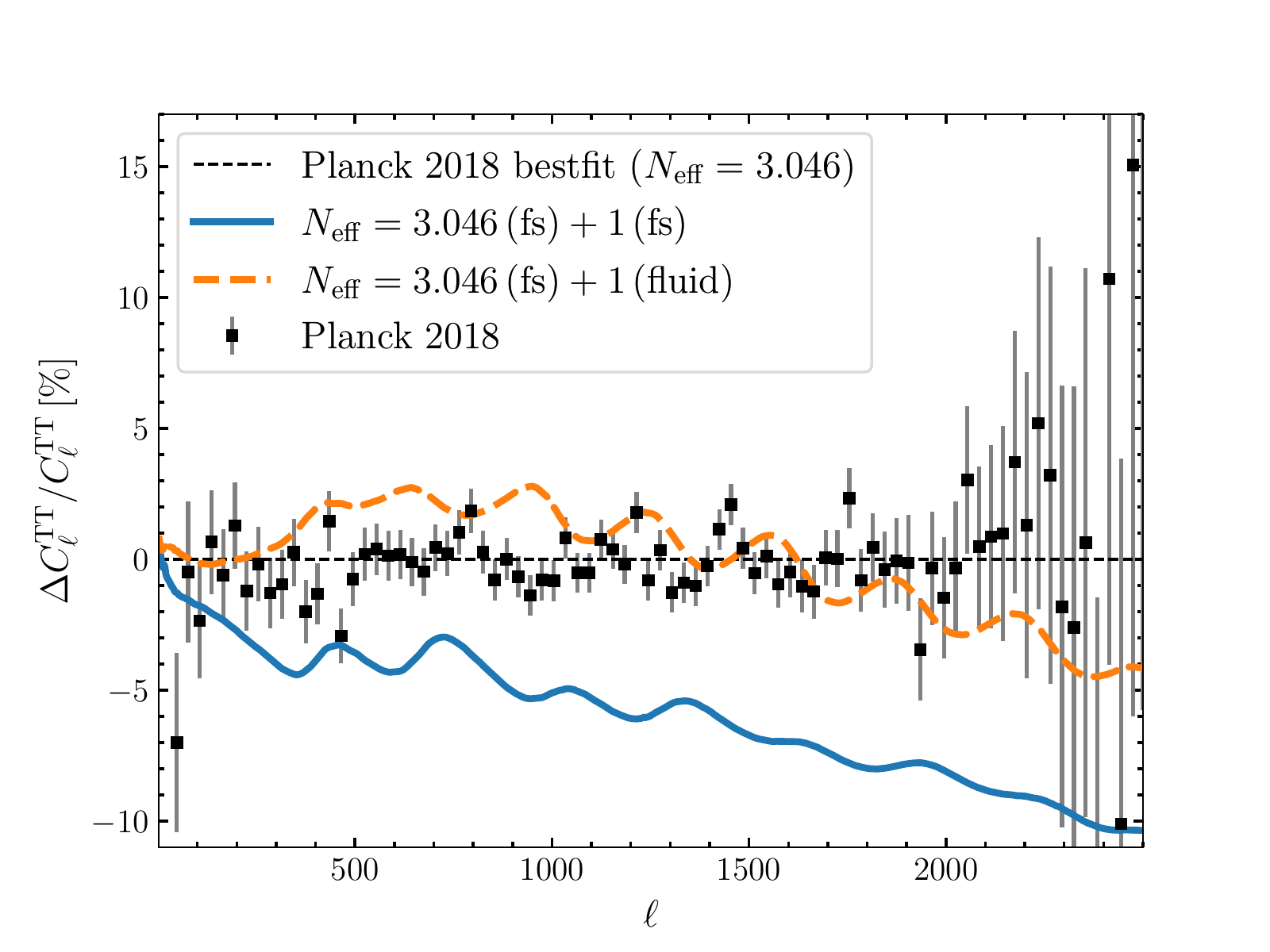}
\caption{
Percentage variation of the CMB temperature power spectrum with respect to Planck 2018 best fit. The spectra are computed fixing $\theta_{\rm s}$, $\Omega_{\rm b}$ and $z_{\rm eq}$, and increasing $\Neff$ by adding one additional species that behaves either like standard neutrinos, i.e.\ free streaming (fs, blue solid line), or like a fluid (orange dashed line). The shift of the acoustic peaks with respect to the reference ($\Neff=3.046$) is out of phase in the two cases. The damping is reduced when the additional species behaves like a fluid. Planck 2018 measurements are shown for reference (black points with error bars). Notice that here the reference value is $\Neff=3.046$, rather than the updated $\Neff=3.044$, for consistency with the best fit provided by the Planck Collaboration.
}
\label{fig:TT}
\end{figure}
It is well known that adding relativistic degrees of freedom reduces the size of the sound horizon at recombination $r_{\rm s}$, thus, increasing $H_0$ and apparently solving the tension.
However, in order to keep the angular scale to the sound horizon $\theta_{\rm s}=r_{\rm s}/D_{\rm A}$ fixed at the value accurately measured by CMB, the angular diameter distance $D_A$ must decrease by the same amount as $r_{\rm s}$.
As a consequence the damping angular scale $\theta_{\rm d}=r_{\rm d}/D_{\rm A}$ increases, because the decrease in the damping scale $r_{\rm d}$ is not as fast as the one in $r_{\rm s}$ (and in $D_{\rm A}$) \cite{Knox:2019rjx} (see Figure~\ref{fig:dist} and Figure~\ref{fig:3D}).
Because of this excess of damping, additional thermal relativistic species are disfavoured by CMB.
Thus, simply varying $\Neff$ has proven to be unsuccessful both for solving the Hubble problem and for accommodating sterile neutrinos in Planck data \cite{Archidiacono:2020yey}.

The impact of additional relativistic species on the CMB changes in the presence of neutrino non-standard interactions. Indeed, as long as the interaction is efficient, neutrino free-streaming is negligible.
The absence of free streaming counteracts the aforementioned background effects of a larger $\Neff$ (e.g.\ the damping excess) by modifying the neutrino behaviour at perturbation level. Indeed the gravitational tug due to free streaming neutrinos leads to a phase shift of the CMB acoustic peaks towards larger scales (smaller $\ell$), and to an overall suppression of their amplitude \cite{Bashinsky:2003tk}. In the absence of free streaming these effects disappear, compensating the background effects due to additional neutrino species (see Figure~\ref{fig:TT}).
Depending on whether the interaction is confined to the sterile sector or extended to the active one, the absence of free streaming affects, respectively, only the additional sterile species (see Figure 4 of Ref.~\cite{Archidiacono:2020yey}), or all neutrinos (see Figure 1 of Ref.~\cite{Kreisch:2019yzn}).

In some models of NSI, the combination of these background and perturbation effects allows not only to accommodate sterile neutrinos in cosmological data, but also to solve the Hubble constant problem \cite{Archidiacono:2020yey,Brinckmann:2020bcn, RoyChoudhury:2020dmd, Blinov:2020hmc,Kreisch:2019yzn,Blinov:2019gcj,Archidiacono:2016kkh}.
Looking at model comparison, the NSI models fit cosmological data better than $\Lambda$CDM only when the prior on the Hubble constant is included in the analysis. On the other hand, the fit of CMB-only data is worse than in $\Lambda$CDM, particularly for Planck E-mode polarization, where the peak structure is less affected by additional effects such as the Doppler effect \cite{Archidiacono:2020yey}. However, replacing Planck data with Atacama Cosmology Telescope (ACT) data yields a negative $\Delta \chi^2$ of the pseudoscalar model with respect to $\Lambda$CDM, pointing to a mild preference of ACT for the pseudoslacar \cite{Corona:2021qxl}.

\begin{figure}[h]
\centering
\includegraphics[trim=0.5cm 0 0 0, width=0.6\textwidth, clip]{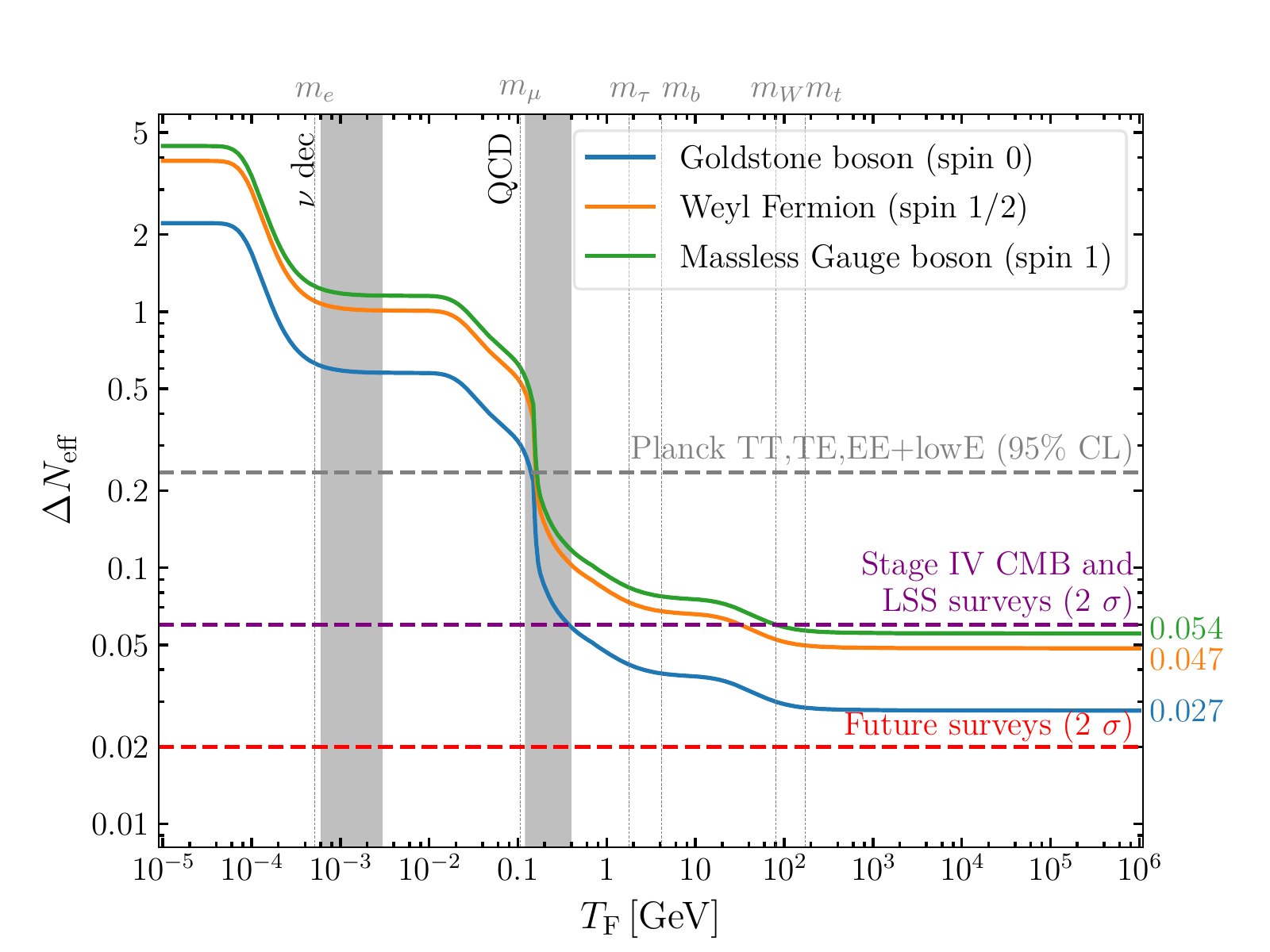}
\caption{Contribution to $\DNeff$ from new light particles with different spin: a Goldstone boson (blue line), a Weyl fermion (orange line), and a massless Gauge boson (green line). Vertical grey rectangles indicate the temperature of QCD phase transition ($\sim 200$ MeV) and the temperature of neutrino decoupling ($\sim 1$ MeV). Current Planck bounds already exclude the existence of particles decoupling after the QCD phase transition. Upcoming CMB and Large Scale Structure (LSS) surveys will improve by two orders of magnitude the limit on the decoupling temperature of new Gauge bosons and of new Weyl fermions. Finally, the sensitivity of future cosmological surveys, including 21 cm surveys, will allow to exclude the presence of new light particles for any decoupling temperature and for any spin.
}
\label{fig:DNeff}
\end{figure}

Future galaxy surveys (Euclid \cite{EUCLID:2011zbd}, Vera Rubin Observatory \cite{LSSTScience:2009jmu}) and CMB surveys (CMB Stage-4 \cite{CMB-S4:2016ple}) will improve the sensitivity to $\Neff$, thus, shedding light on scenarios, such as NSI, that require new light mediators \cite{Park:2019ibn}.
As shown in Equation \ref{eq:Tdec}, the contribution of new particles to $\Neff$ depends on their freeze out temperature $T_{\rm F}$.
Current bounds on $\DNeff$ ($\DNeff \lesssim 0.3$ \cite{Planck:2018vyg}) already rule out particles that decoupled below the QCD phase transition $T_{\rm F} \sim 200$~MeV.
The sensitivity of Euclid to $\Neff$ can reach $\sigma(\Neff)=0.046$ \cite{Sprenger:2018tdb} in combination with Planck, and $\sigma(\Neff)=0.03$ \cite{Sprenger:2018tdb} in combination with CMB S4 \cite{Baumann:2017gkg,Brinckmann:2018owf}; finally the inclusion of future 21 cm intensity mapping can further improve the sensitivity to $\sigma(\Neff)=0.01$ \cite{Castorina:2020zhz, Sailer:2021yzm} (see Figure \ref{fig:DNeff}).
Notice that the minimum value of $\DNeff$ allowing for the presence of new light particles is $0.027$, i.e.\ the asymptotic value at high temperature for a scalar boson.
Therefore, future limits on $\Neff$ will open the window on the thermodynamics of the Universe back to the reheating temperature.
In the near future, many cosmological models entailing new particles in the form of dark radiation might be excluded with high significance.

\begin{figure}[h]
\centering
\includegraphics[width=0.5\textwidth]{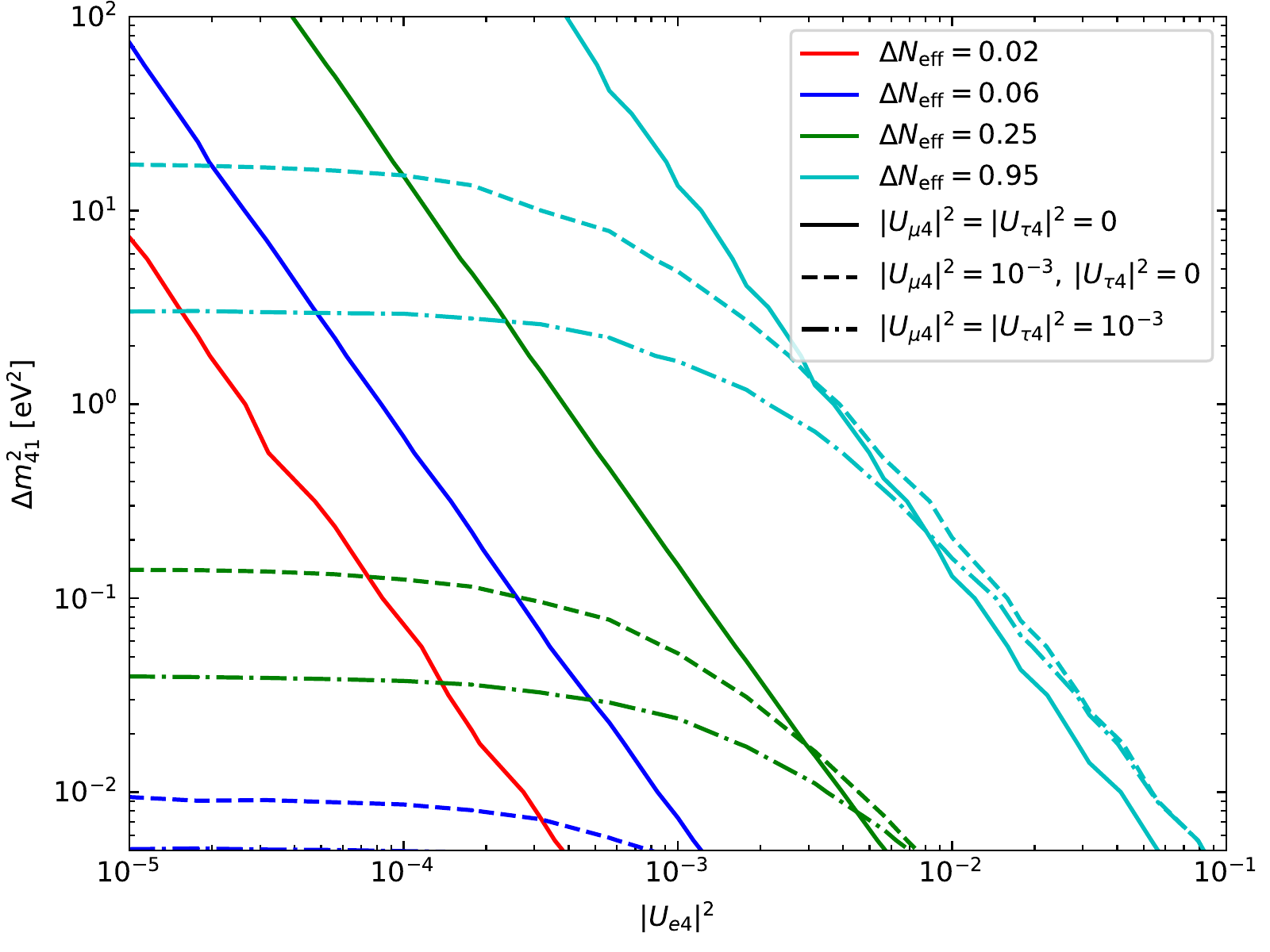}
\caption{Sterile neutrino contribution $\DNeff$ as a function of the active-sterile mixing parameters, for different values of the mixing matrix elements. Colors represent different levels of $\Neff$ as follows: 
$\DNeff=0.02$ (red, $2\sigma$ future surveys \cite{Castorina:2020zhz, Sailer:2021yzm}),
$\DNeff=0.06$ (blue, $2\sigma$ stage IV CMB and LSS surveys \cite{Sprenger:2018tdb}),
$\DNeff=0.25$ (green, 95\% CL limit from Planck TT,TE,EE+lowE+lensing+BAO, for $m_s<10$~eV \cite{Planck:2018vyg}),
and $\DNeff=0.95$ (i.e.\ $\Neff\sim4$, cyan, practically fully thermalised sterile neutrino).
}
\label{fig:Neff_mixpar}
\end{figure}

Concerning the sterile neutrino, the $2\sigma$ limits mentioned above may also be translated into constraints on active-sterile mixing parameters.
Figure~\ref{fig:Neff_mixpar} shows the dependence of the sterile neutrino contribution to $\Neff$, $\DNeff$, on the new mass splitting \dm{41} and mixing matrix elements, in particular \Usq{e4} varies in the plot, while different combinations of fixed $\Usq{\mu4}$ and $\Usq{\tau4}$ correspond to different line styles as indicated in the legend.
Colored lines represent different iso-$\DNeff$ contours:
mixing parameters corresponding to a fully thermalised state correspond to lie in the region to the right of the cyan lines ($\Neff\sim4$);
the current 95\% CL limit from Planck TT,TE,EE+lowE+lensing+BAO \cite{Planck:2018vyg} (obtained marginalising over $m_s<10$~eV) is shown in green;
finally, blue and red lines represent future $2\sigma$ sensitivities from future stage IV CMB and LSS surveys, and from future 21~cm surveys, respectively.
Notice that these latter probes may correspond to an order of magnitude stronger constraints on the mixing matrix elements between active and sterile states with respect to current bounds.

\section{Interacting dark radiation}
\label{sec:darksector}

Dark radiation, sterile neutrinos, and non standard interactions, can be seen as different aspects of a more complex dark sector, including also dark matter. Interactions between dark radiation and dark matter are motivated not only by the cosmological tensions, i.e.\ the $H_0$ problem and the $\sigma_8$ tension \footnote{The $\sigma_8$ tension refers to the $\sim 2 \sigma$ deviation between CMB observations and weak gravitational lensing data on the clustering of matter at scales of $8\,h/$Mpc \cite{DES:2020hen}.}, but also by the lack of detection of dark matter in the form of Weakly Interacting Massive Particles, and by the debatable mismatch between Cold Dark Matter predictions and observations on very small scale \footnote{About the so-called Cold Dark Matter small scale crisis \cite{Bullock:2017xww} notice that, at the state of the art, baryonic feedback can solve some of the problems \cite{Burger:2021sep,Governato:2012fa,Zolotov:2012xd}, but there is no consensus yet on the ability of baryons to solve all the problems \cite{Oman:2015xda}.}.

\begin{figure}[!h]
\centering
\includegraphics[trim=0.4cm 0 0 0, width=0.5\textwidth, clip]{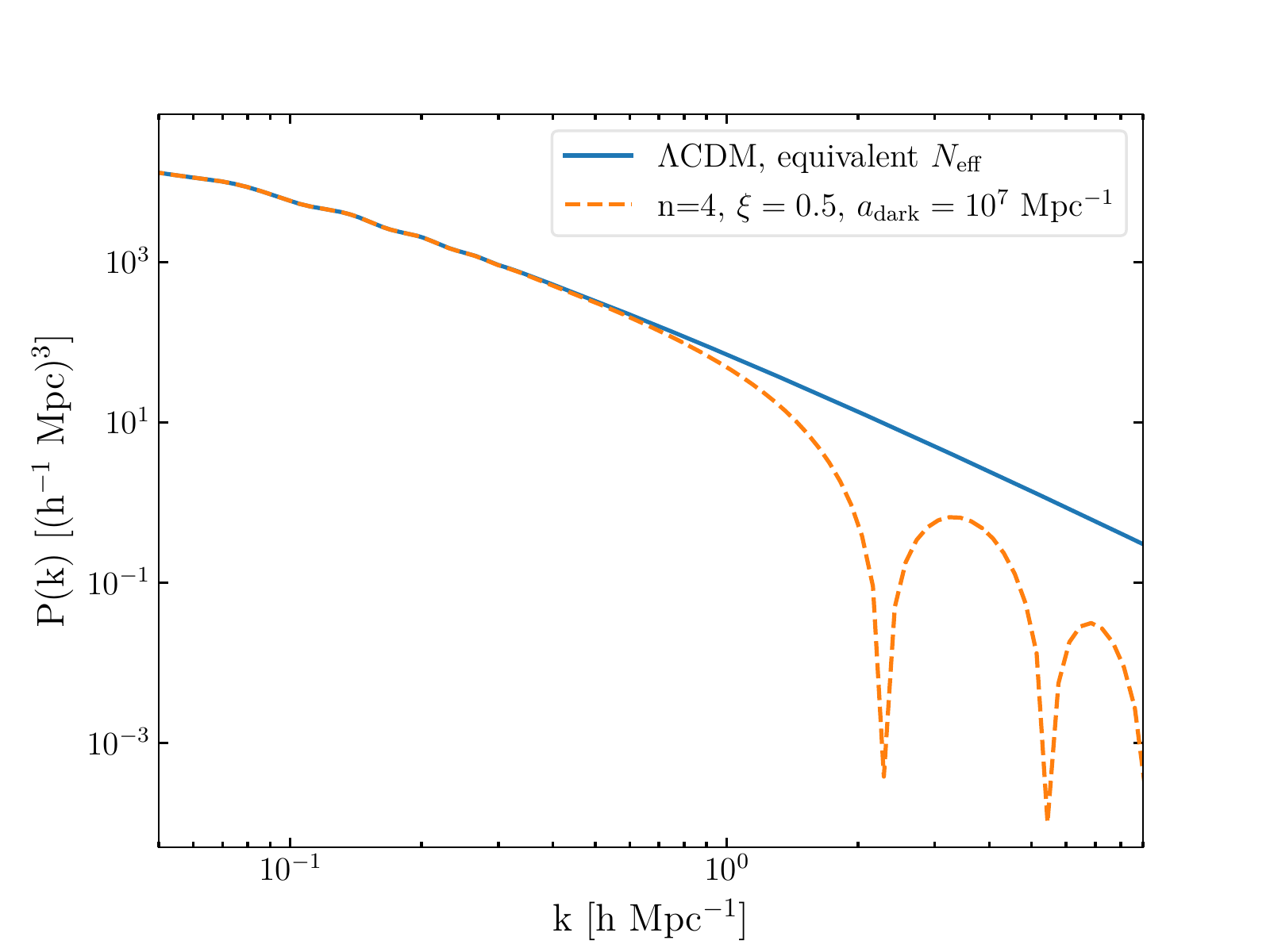}
\caption{
Matter power spectrum for $\Lambda$CDM (blue solid line) with the same $\Neff$ as in the interacting dark matter-dark radiation case (orange dashed line) where the amount of dark radiation is defined by the temperature ratio $\xi=T_{\rm DR}/T_\gamma=0.5$. The index $n$ denotes the temperature dependence of the comoving interaction rate, while the parameter $a_{\rm dark}$ represents the strength of the interaction; in order to show the impact of the interactions we chose a value of $a_{\rm dark}$ much larger than the current limits. The plot is made with the \texttt{CLASS} version publicly released with Ref.~\cite{Archidiacono:2019wdp}.
}
\label{fig:n4}
\end{figure}

Recently, several particle models of dark matter -- dark radiation interactions were proposed and tested in the literature \cite{Cyr-Racine:2013fsa, Chu:2014lja, Buen-Abad:2015ova, Lesgourgues:2015wza, Cyr-Racine:2015ihg, Schewtschenko:2015rno, Krall:2017xcw, Archidiacono:2017slj, Buen-Abad:2017gxg, Archidiacono:2019wdp, Becker:2020hzj}.
The variety of the underlying microphysics of the different models (e.g., non-Abelian dark matter \cite{Buen-Abad:2015ova, Lesgourgues:2015wza, Buen-Abad:2017gxg, Krall:2017xcw}, atomic dark matter \cite{Cyr-Racine:2013fsa}, dark matter interactions mediated by new bosons \cite{Chu:2014lja}) can be mapped into few phenomenological parameters (ETHOS \cite{Cyr-Racine:2015ihg, Bohr:2020yoe}), and categorised according to the temperature dependence of the scattering rate ($\Gamma \propto T^n$).
When $\Gamma>H$, being $H$ the Hubble rate, the drag force of dark radiation on dark matter is efficient and prevents dark matter clustering. The delay of clustering after the decoupling of the interaction ensures a suppression of power on small scales, and thus a possible solution to the Cold Dark Matter small scale crisis.
As an example, we show in Figure~\ref{fig:n4} (orange dashed line) the power spectrum of dark matter-dark radiation interactions with a temperature dependence $n=4$ of the comoving interaction rate. This would be the case of dark matter particles interacting with light sterile neutrinos through a new massive boson. We notice a series of dark acoustic oscillations, due to the relativistic pressure counteracting the gravitational collapse \cite{Buckley:2014hja}, and an exponential suppression with respect to the equivalent $\Lambda$CDM spectrum (blue solid line).
Concerning the cosmological tensions, only models where the interaction rate has the same temperature dependence of the expansion rate seem to alleviate the $H_0$ problem and the $\sigma_8$ discrepancy \cite{Archidiacono:2019wdp,Becker:2020hzj}, although the choice of priors on the interaction parameters might nullify this possibility \cite{Diacoumis:2018ezi}.

Bounds from linear scales (CMB) and mildly non-linear scales (Baryonic Acoustic Oscillations) have greatly reduced the allowed range of values of the interaction strength.
However, the most relevant features of these models are located at very small non-linear scales, such as those probed by Lyman-$\alpha$ data \cite{Archidiacono:2019wdp}. Therefore, future data constraining the matter power spectrum on small scales, and a better understanding of the modelling of the non-linear clustering on those scales, will be crucial to further constrain models of dark matter -- dark radiation interactions.

\section{Conclusions}
\label{sec:conclusions}

In this article the case of light sterile neutrinos, favoured by some neutrino oscillation anomalies, has been reviewed in connection with their cosmological interpretation in terms of dark radiation.

Concerning neutrino oscillation anomalies, several experiments are currently taking data, either using neutrinos from nuclear reactors, radioactive sources, accelerated beams or interactions in the atmosphere.
Each set of experiments constrains a different mixing channel and only their combination can provide global information on active-sterile neutrino mixing parameters.
Unfortunately, several disagreements are found between different experiments or classes of probes.
The most statistically significant tension emerges between appearance (mostly by LSND and MiniBooNE) and disappearance (primarily reactor and Gallium experiments, IceCube, MINOS/MINOS+) observations,
which have been incompatible with one another since many years.
Such strong tension is under investigation by several planned or ongoing experiments, such as MicroBooNE, whose recent results suggested that the MiniBooNE excess may not be due to neutrino oscillations, therefore disfavoring the existence of a sterile neutrino.
In the coming years, the data we expect to receive from these experiments will help us to understand better current results, potentially either confirming or excluding the sterile neutrino hypothesis.

Besides oscillation experiments, sterile neutrinos would have an impact on several other observables.
From the terrestrial point of view, we discuss the constraints obtained from $\beta$ decay kynematics and neutrinoless double $\beta$ decay probes.

When we turn to cosmology, eV sterile neutrinos would be too many ($\Neff \sim 4$) and too heavy ($\summnu \sim 1$ eV) to fit cosmological data both at high redshift (Cosmic Microwave Background, CMB) and at low redshift (Large Scale Structure, LSS), if one assumes the standard production mechanism through neutrino oscillations.

If future short baseline experiments will confirm the presence of sterile neutrinos, the inconsistency with cosmology could be solved by means of neutrino non standard interactions (NSI). NSI prevent thermalisation, thus lowering the contribution of sterile neutrinos to $\Neff$, and making it consistent with the CMB bounds. Moreover, the pseudoscalar case of NSI can also prevent the contribution of the sterile neutrino mass to the hot dark matter density of the Universe, thus the LSS bounds on $\summnu$ do not apply. At the same time, NSI can also alleviate the Hubble constant problem by modifying the neutrino behaviour at perturbation level: indeed, in the presence of NSI, neutrinos do not free stream.

New interactions require the presence of a new mediator, which also represents a form of dark radiation contributing to $\Neff$. Current limits on $\DNeff$ still leave room for the presence of these new particles, of course depending on their thermal history. However, the endgame for a new mediator, and, thus, for new interactions, might be within reach of the sensitivity of upcoming and future cosmological surveys.

\authorcontributions{All the authors contributed equally to this work at all stages. All authors have read and agreed to the published version of the manuscript.}

\funding{
SG acknowledges financial support from the European Union's Horizon 2020 research and innovation programme under the Marie Skłodowska-Curie grant agreement No 754496 (project FELLINI).
}

\acknowledgments{In this section you can acknowledge any support given which is not covered by the author contribution or funding sections. This may include administrative and technical support, or donations in kind (e.g., materials used for experiments).}

\conflictsofinterest{The authors declare no conflict of interest. The funders had no role in the design of the study; in the collection, analyses, or interpretation of data; in the writing of the manuscript, or in the decision to publish the~results.} 


\reftitle{References}


\externalbibliography{yes}
\bibliography{bibliography}

\end{paracol}
\end{document}